\documentclass{aa}
\usepackage[dvips]{graphicx}
\usepackage{amssymb}
\usepackage[dvips]{graphicx}
\usepackage{graphics}
\begin{document}

\title{ Polarization statistics of extra-solar systems }
\author{
F. Tamburini\inst{1}
\and
S. Ortolani\inst{2}
\and A.
Bianchini\inst{2} }
%
%
\institute{
Institute of Cosmology and Gravitation, Univ. of
Portsmouth, PO2 1EG, Portsmouth, UK\\
email: fabrizio.tamburini@port.ac.uk
\and
Universit\`a di Padova, Dipartimento di Astronomia, Vicolo
dell'Osservatorio 2, I-35122 Padova, Italy\\
emails:  bianchini@pd.astro.it, ortolani@pd.astro.it\\
}
\date{Received ; accepted }
\abstract{We have analysed the optical polarimetric properties of
nearby intermediate-late type MS stars and examined in some detail
the pecularities displayed by those known to have planets. We find
that {\it i}) the polarization degree is not strictly correlated
with the presence of planets; {\it ii}) there is a lack of high
eccentricity planets at high  optical linear polarization levels;
{\it iii}) no clear correlation is seen between metallicity and
polarization amongst extra-solar systems; {\it iv}) the
contribution to the polarization  by the interstellar medium seems
to become effective only after $\sim 70~ pc$.  }
\titlerunning{Polarization of extra-solar systems}
\authorrunning{F. Tamburini et al.}
\maketitle
%

\section{Introduction}
The search for extra-solar planets represents an exciting new
frontier for modern astronomy. The results obtained with Doppler
surveys are still strongly limited both by the sensitivity of the
detectors and by the long recording of observations in time
needed. This selection effect seems to be the main reason why most
of the planets within a few parsecs of the Sun are massive and
fast revolving in very close orbits around their central stars.
Nevertheless, we can try to identify some general characteristics
by analyzing the statistical properties of the available data.

The statistical properties of extra-solar systems show that stars
harbouring massive planets display  metal-rich features with
respect to  nearby field stars (Reid 2001, Santos et al. 2001). At
sufficient distances from the star, the decreasing  temperature of
the circumstellar disks allows for the presence of $\sim 1 - 2\%$
of  solid particles, interstellar grains or condensates, that can
be detected from polarization measurements (Voshchinnikov \&
Kr\"ugel  1999) as, for example, in $\beta$  Pictoris.

Investigation of our solar system showed that the dominant source
of dust inside Jupiter's orbit is represented by short-period
comets. Outside Jupiter's orbit, the concentration of particles is
a function of the distance from the Sun and the material moves
along randomly eccentric orbits (Humes 1980). Landgraf et al.
(2002) showed that the dust between Jupiter and Saturn is still
mainly produced by comets, while, outside Saturn's orbit, it is
mainly formed of micrometer-sized grains produced by  mutual
collisions  of the Edgeworth-Kuiper Belt objects and also by the
impact of interstellar dust grains. Data from  Voyager 1 and 2
spacecrafts show also the existence of a high concentration of
particles beyond $50~ au$ (Gurnett et al. 1997).

These results boosted the hunt for planets around distant stars
through the detection of circumstellar dusty disks. These disks
can produce observable effects such as infrared emission excess
(Beckwith and Sargent, 1996; Spangler et al., 2001) and partial
polarization of the star light reflected via scattering
(Mauron and Dole, 1998; Yudin, 2000). Some polarization
could also be produced by a giant planet in a close orbit around
the star (Saeger et al. 2000), but the resulting effect is too
weak to be responsible for the polarizations actually observed.

In this paper, we investigate the statistical properties of the
linear polarization of  a sample of extrasolar systems and compare
them with those of  nearby stars. In section 2 we report the
statistical properties of our data sample related to the
polarization in white light. In section 3  we draw our
conclusions.

\section{Polarization properties}

The polarization  data are taken from the Heiles catalogue
(Heiles 2000) by selecting stars within a distance of $70~ pc$,
that roughly corresponds to the largest distance amongst the
observed extrasolar planet systems so far observed. The main
characteristics of the extra-solar planetary systems within our
sample are taken from the Planet Encyclopedia (Schneider 2002).
The errors in the polarization measurements are those reported in
Heiles catalogue, and whenever a star is listed in more than one
catalogue, the polarization values are given as weighted
averages. The typical error of the polarization degree at the $1
\sigma$ level  ranges from $0.032\%$ to $0.1\%$. Very uncertain
data are excluded in our sample of planetary systems.

To analyse the statistical polarization properties of the observed
extrasolar systems, we have compared their polarization
distribution to those of stars of our sample that are not known to
host planets. Since we do not observe extrasolar systems with
$p>0.09$, we have taken this value  as an upper limit  also for
the sample of single stars. We notice in Fig. 1 that the two
distributions are different as confirmed by the Kolmogorov-Smirnov
test at the $99\%$  confidence level. Although both the
distributions show a peak at low polarization followed by a
decrease towards larger values, it appears that single stars show
a maximum between $p=0.01-0.02$, while stars with planets  show a
well defined maximum around $p\sim0.0$ and a rather steeper
decline towards higher polarization degrees. This seems indeed a
quite surprising result.

The general negative trend at increasing polarization might appear
reasonable for both the samples, but why should extrasolar systems
cluster around zero?

\begin{figure}[!htb]
\includegraphics[clip=,width=0.5\textwidth]{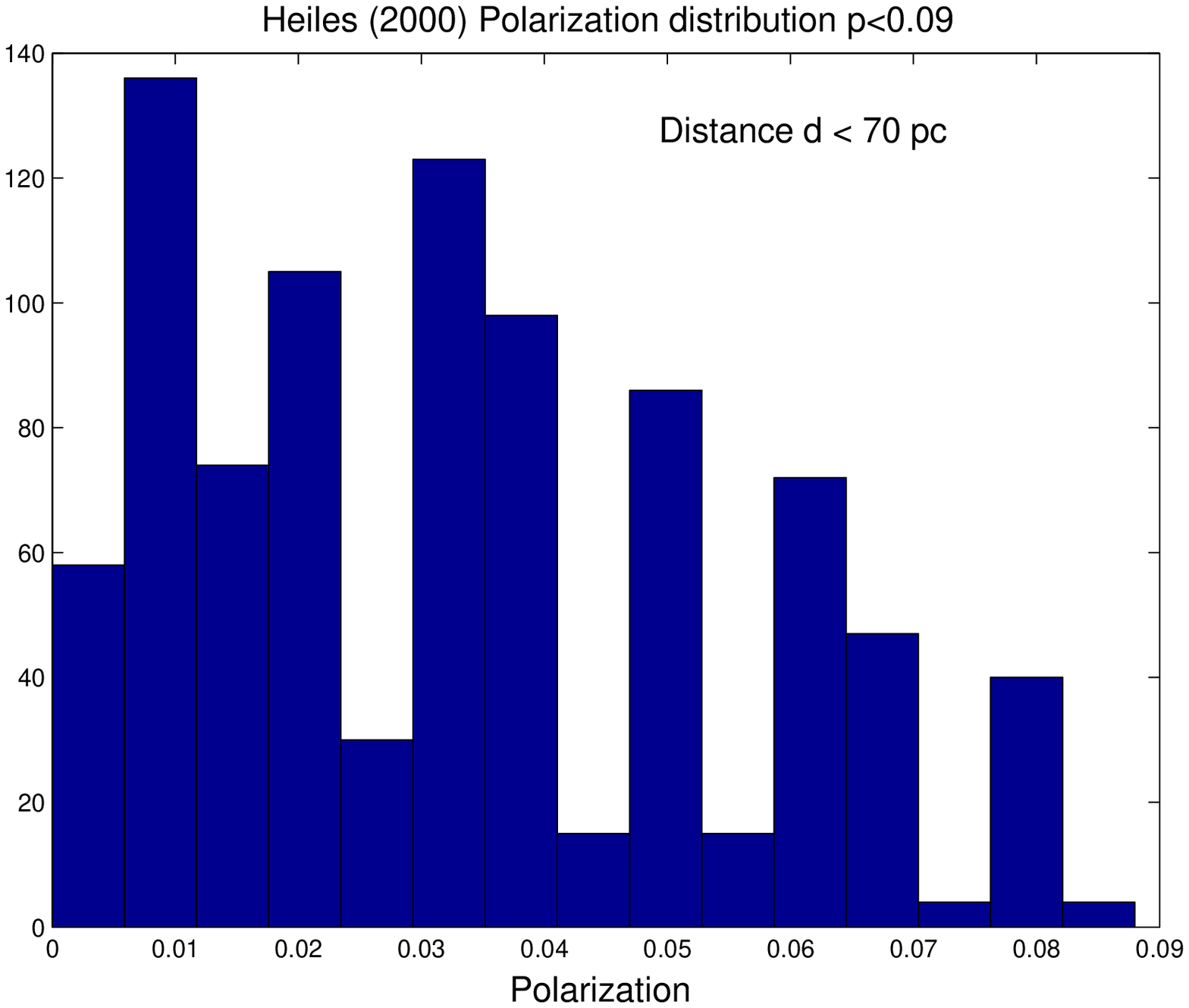}
\includegraphics[clip=,width=0.5\textwidth]{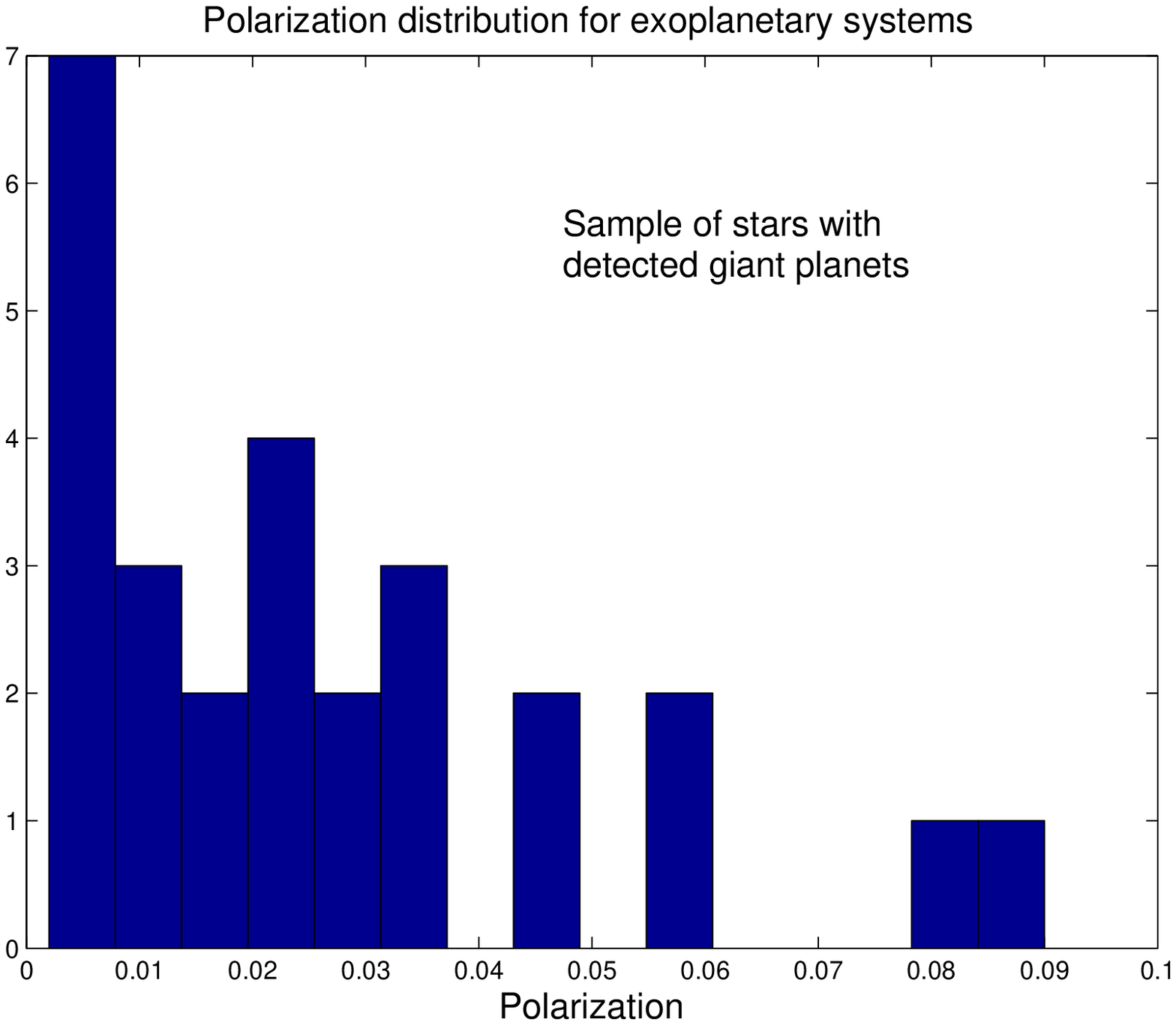}
\caption[]{({\bf Upper panel}) Histogram of the distribution of
the polarization within $70 pc$ and $p <0.09$  of stars from
Heiles (2000) Catalogue within $70 pc$ and $p <0.09$.\\ \noindent
({\bf Lower panel}) Polarization distribution  of the sample of
stars with detected giant planets. A well defined peak around
$p\sim 0$ is evident.\\ \noindent According to the
Kolmogorov-Smirnov test the two distributions are different.}
\label{pol1}
\end{figure}

\begin{figure}[!htb]
\includegraphics[clip=,width=0.5\textwidth]{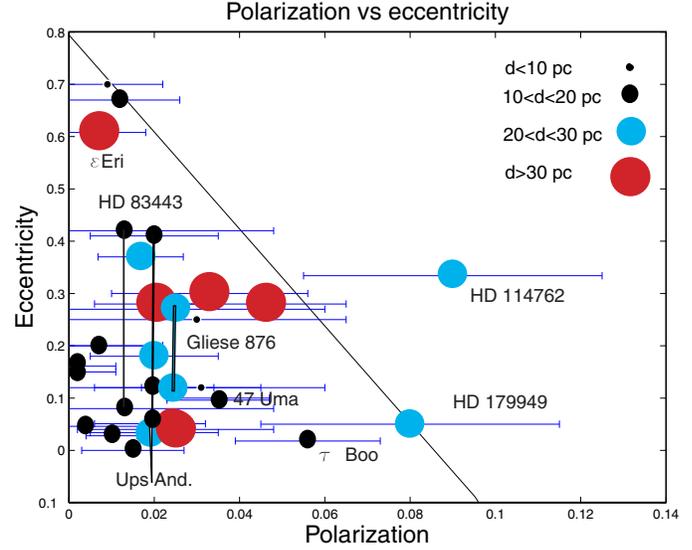}
\caption[]{Polarization $vs$ orbital eccentricity. The extrasolar
systems are mainly distributed in the lower
left region of the diagram.  The diagonal line tentatively separates the
two regions in the diagram. The exception of HD 114762 is
discussed in the text. There is indication that larger polarizations
are allowed at lower eccentricities. Multiple
systems reported in Table 1 are shown connected by dotted lines.
Symbol sizes correspond to different distances: no distance
effects can be seen.} \label{polecc}
\end{figure}

Plotting in Fig. 2 the orbital eccentricities of the extrasolar
planets $vs$ the observed polarization, we find a peculiar
distribution of the data points, since only the  lower left part
of the diagram is populated, while the upper right region is
almost empty. Given this surprising result, we investigated
possible systematic effects. Distance effects seem to be
irrelevant for the observed distribution because distances are
randomly distributed with respect to the polarization. This is
confirmed in  Fig. 3 where the polarization of FGK stars is
plotted against the distances. We see that within the distance of
about 70 pc which corresponds to the limit of our sample, no
correlation is seen. Considering that the mass of the disks have
been suggested to be a function of their age (Spangler et al.,
2001), we checked if the observed higher polarization of HD 179949
and Tau Bootis could be due to a younger age. HD 179949 is 3.31
Gyr old, and Tau Bootis is 2.52 Gyr (Chen, 2001; Marsakov, 1995),
while, for example Epsilon Eridani, which has a rather low
polarization ($p=0.01$), is instead rather younger (about 1 Gyr).
Figure 2 suggests that  highly eccentric orbits can only have low
polarization levels, while high polarizations are detected only in
low eccentricity systems. In any case the polarization levels of
many of the stars with planetary systems plotted in the figure
appear surprisingly higher than predicted by both current
theoretical models and observational evidence.

The extra-solar systems can then be compared to $\beta$ Pic, in
which the polarization is rather high, but in our case the dusty
discs around them are very rarefied and no polarization above
$\sim$ $p=0.01$ has so far been found (Spangler et al., 2001).

The data used (Heiles catalogue) are very uncertain: the error
bars in the polarization degree are often larger than the measured
polarization, but at low eccentricity values they are
significantly smaller than the data dispersion.

At present it is not clear what the mechanism could be for the
observed polarization levels in our sample. If the trend of the
diagram in Fig. 2 is real and the disks are not responsible for
the whole observed polarization, we should look for some other
polarization mechanisms somehow connected to the eccentricities.

We note that the F9V star HD 114762 system, which is the
only exception, presents
$p=0.09$ and  $e=0.334$ but it is also peculiar because
its companion is a substellar object, located at only $ 0.35~ au$ from
the star, having a mass $m \sin i=0.011\pm
0.001 M_{\odot}$, that is just above the limit for H fusion.
Thus, this object
should be  considered more as a brown dwarf candidate
(Latham, et al. 1989) rather than a  planet.

A consequence of Fig. 2 is that the peak at $p\sim0.0$
in the bottom histogram  of  Fig. 1 might  be caused by the
presence of highly eccentric systems.

\begin{figure}[!htb]
\includegraphics[clip=,width=0.5\textwidth]{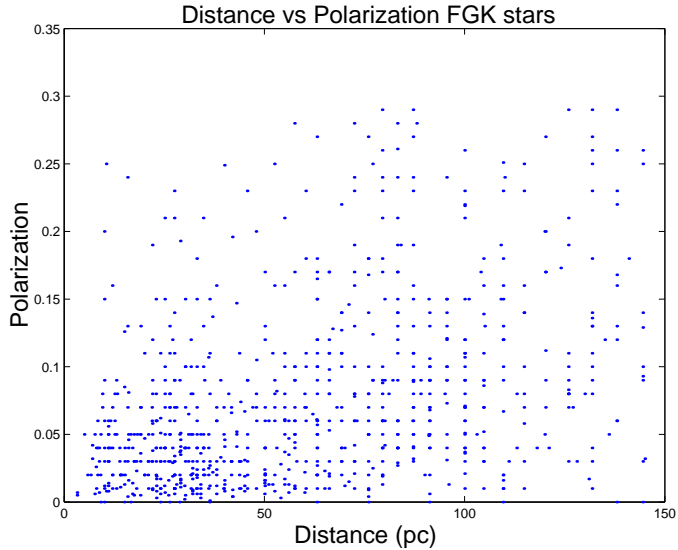}
\caption[]{Polarization $vs$ distance from the Sun of FGK stars.
No correlation is seen within the limit distance of 70 pc of our
sample (see text).} \label{poldist}
\end{figure}

We have also looked for possible statistical  effects on the
polarization due to metallicity. The plot of Fig. 4 confirms the
known  larger metallicities observed in extrasolar systems as
compared to the Sun (Boss 2002) and shows  quite a large
dispersion with no clear evidence of systematic trends. However,
though not statistically significant, we notice that the two
systems HD 6434 and HD 114762, which present the lowest
metallicities, [Fe/H]=-0.55 and [Fe/H]=-0.79, respectively, also
display relatively large polarizations, giving the impression of a
negative correlation.

\begin{figure}[!htb]
\includegraphics[clip=,width=0.5\textwidth]{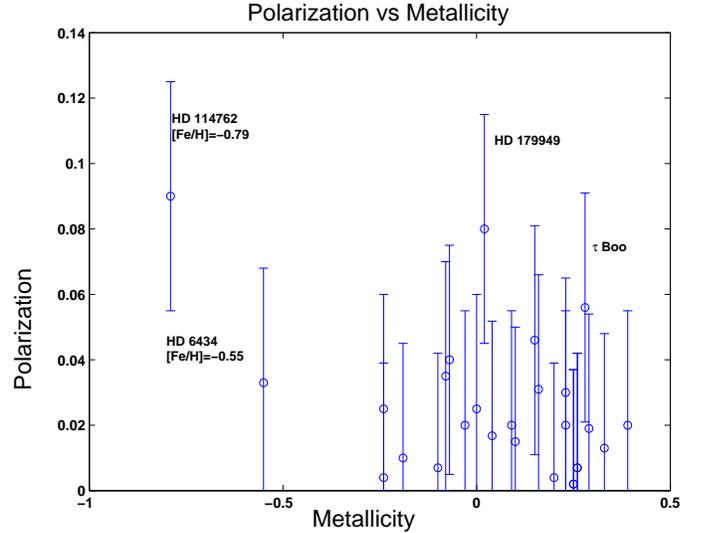}
\caption[]{Polarization $vs$ metallicity. No clear correlation is
seen. Planetary systems HD 6434 and HD 114762 might be peculiar.
} \label{polmet}
\end{figure}

The characteristics of the only four multiple planetary systems,
whose polarization is known, are given in Table 1; p is the degree
of polarization, Tsp is the spectral type of the host star, a is
the semi-major axis of the orbit, P is the orbital period in days
and e the eccentricity

The eccentricities of the planets in each system are represented
in the plot of Fig. 2 connected by dotted lines. These systems
show a medium-low polarization degree, between the value $p=0.013$
of HD 83443, in which one of the planets presents the largest
eccentricity  $e \sim 0.4$, and the value $p=0.035$ of 47 UMa,
which has two planets with almost the same low eccentricity $e\sim
0.1$. This seems to be consistent with the general behavior
displayed in Fig. 2 in that larger eccentricities correspond to
lower polarization levels. However, for the moment, no further
speculation can be made until more observations of multiple
systems  are collected.

\section{Conclusions}

We have analyzed the statistical polarimetric properties of
extrasolar systems taken from  the Heiles catalogue (Heiles 2000).
We have compared the polarization distribution of extrasolar
systems with that of a selected sample of stars with no planets
detected. The two distributions might be similar if we remove a
few high eccentricity extrasolar systems that presents almost zero
polarization.
\smallskip

The most significant result  found is  that only systems with low
eccentricity orbits can have high polarization values.

No correlation between polarization and metallicity is evident. We
have also found that at distances $< 70 ~pc$  no relevant
contribution by interstellar dust is expected.

We have started an observational spectropolarimetric  program at
the Asiago Observatory to analyse separately the polarization
produced by the interstellar dust and that produced by
circumstellar material with a higher accuracy and for a wider
sample of extrasolar systems. Preliminary results show that the
polarization accuracy we can achieve is of the order of $0.01\%$
and the low polarization of the high eccentricity systems seems
confirmed.

\begin{acknowledgements}
We acknowledge Bruce Bassett for helpful discussion.
We are very grateful to Nikolai Voshchinnikov for his helpful comments.
The present work has been supported by the italian Ministero della
Universit\`a e della Ricerca. FT whishes to thank the Department
of Astronomy of the University of Padova for the kind hospitality.

\end{acknowledgements}

\begin{table}[h]
\caption{\label{Tab1} Properties of multiple extrasolar systems.}
\begin{tabular}{ccccccc}
Name &  p   & Tsp & m sin i & a (au) & P (d) & e\\ \hline HD83443a
& 0.013 & K0V & 0.35 & 0.038 & 2.9861 & 0.08\\ HD83443b & 0.013 &
K0V & 0.16 & 0.174 & 29.83 & 0.42\\ Ups And a & 0.02 & F8V & 0.71
& 0.059 & 4.6170 & 0.034\\ Ups And b & 0.02 & F8V & 2.11 & 0.83 &
241.2 & 0.18\\ Ups And c & 0.02 & F8V & 4.61 & 2.50 & 1266.6 &
0.41\\ Gliese 876a & 0.025 & M5 & 0.56 & 0.13 & 30.1 & 0.12\\
Gliese 876b & 0.025 & M5 & 1.98 & 0.21 & 61.02 & 0.27\\ 47 Uma a &
0.035 & G0V & 2.41 & 2.1 & 1095 & 0.096\\ 47 Uma b & 0.035 & G0V &
0.76 & 3.73 & 2594 & 0.1 \\ \hline
\end{tabular}
\end{table}


%
%

%
\end{document}